\documentclass{article}
\usepackage{a4}
\usepackage{epsfig}
\usepackage{lineno,url}
\newcommand\eps{\varepsilon}

\newcommand\pt{\partial}
\begin{document}
\title{Beam Optics Primer\\ using Octave or MATLAB}
\author{V. Ziemann, Uppsala University}
\date{June 16, 2019}
\maketitle
\begin{abstract}\noindent
This primer provides a basic introduction to beam optics concepts that are
commonly used to describe charged particle accelerators. 
\end{abstract}
%
%
\section{Introduction}
The purpose of this document is to prepare the reader for the tutorial sessions
on the forthcoming CERN Accelerator School (CAS), where basic methods for beam 
optics calculations will be illustrated with the help of Octave\cite{OCTAVE}, 
MATLAB~\cite{MATLAB}, or Python~\cite{PYTHON}. At the school, the students 
will be asked to ``play around'' with the software and then
use it to solve simple beam optics problems. This initial familiarization with the 
basic concepts and the software helps to maximize the benefit drawn from the tutorial 
sessions at CAS. Note that this report addresses the absolute newcomer. 
\par
We start by introducing some basic theoretical concepts first, but illustrate the
ideas with, often trivial, examples in Octave as we go along. Exercises are 
interspersed in the text and readers, especially newcomers to the field, are 
encouraged to solve them. We will discuss the solutions at CAS. 
\par
This document focuses on Octave/MATLAB, loosely following~\cite{VZAPB}, but a 
version of this report~\cite{TUTPYTH}, in which all examples are based on 
Python~\cite{PYTHON}, will be available. Most examples in the earlier parts 
of this tutorial are based on one transverse dimension.
\section{Ray tracing}
The motion of charged particles with respect to the center of the beam pipe in
an accelerator conceptually resembles the motion of optical rays with respect 
to the optical axis. In the latter case, a ray at a given longitudinal position 
$s$ is characterized by its distance
$x$ and its angle $x'$ with respect to the optical axis. In the absence of 
lenses or prisms, the light ray (I normally visualize the ray from a laser
pointer) moves on a straight line. Its distance to the optical axis $\hat x$ at
a downstream location $\hat s$ therefore changes according to $\hat x=x+Lx',$
where we assume that $L$ is the distance between $s,$ where the ray's
coordinates are $(x, x'),$ and $\hat s,$ where the coordinates are $(\hat x, \hat x').$ 
Since it moves on a straight line, the angle does not change from $s$ to 
$\hat s,$ such that we have $\hat x'=x'.$ The two equations for $\hat x$ 
and $\hat x'$ can be combined in a single equation,
\begin{equation}\label{eq:drift}
\left(\begin{array}{c} \hat  x\\ \hat x'\end{array}\right)
=\left(\begin{array}{cc} 1 &L\\ 0 & 1\end{array}\right)
\left(\begin{array}{c} x\\x'\end{array}\right)\ ,
\end{equation}
where the motion from $s$ to $\hat s$ is encapsulated in the matrix. 
\begin{itemize}
\item{\bf Exercise 1:} Show that multiplying two such matrices, one with $L_1$ 
  and the other with $L_2$ in the upper right corner, produces a matrix with
  the sum of the distances in the upper right corner.
\end{itemize}
When a ray passes a focusing (thin) lens, its angle $x'$ changes proportional
to the transverse distance $x$ of the ray from the center of the lens, which we assume 
to be aligned with the optical axis. Recall the definition of the {\em focal length}, which 
requires a lens to deflect all parallel rays, independent of their transverse position,
in a such a way, that the rays cross the optical axis at the same point---the focal
point. This point is located a distance $f,$ the {\em focal length,} downstream of the lens.
We therefore require that the change of angle in the lens is given by $\hat x'=x'-x/f.$
Since we assumed the lens to be very thin, the transverse position of the ray does
not change and we have $\hat x=x.$ Here we assumed that the coordinates of the rays,
immediately upstream of the lens are $(x,x')$ and immediately downstream they
are $(\hat x, \hat x').$ Note that also these two equations can be combined into 
the following matrix-valued equation
\begin{equation}\label{eq:quad}
\left(\begin{array}{c} \hat x\\ \hat x'\end{array}\right)
=\left(\begin{array}{cc} 1 &0\\ -1/f & 1\end{array}\right)
\left(\begin{array}{c} x\\x'\end{array}\right)\ ,
\end{equation}
where a $f>0$ describes a focusing lens and $f<0$ a defocusing lens.
\begin{itemize}
\item{\bf Exercise 2:} How do you describe a ray that is parallel to the optical axis?
\item{\bf Exercise 3:} How do you describe a ray that is on the optical axis?
\item{\bf Exercise 4:} Show by multiplying the respective matrices that a parallel ray, 
  which first passes through a lens with focal length $f$ and then moves on a straight 
  line, actually crosses the optical axis at a distance $L=f$ downstream of the lens. 
  Hint: think a little extra about ordering of the matrices!
\end{itemize}
A preliminary analysis of optical systems can easily be described by the matrices,
they are called Ray- or ABCD-matrices in the optics literature~\cite{SIEGMAN}. But
optics is not exactly our topic, and we will therefore turn to charged particles,
instead.
\par
As a matter of fact, we can describe the motion of charged particles in the 
presence of magnetic lenses by the same mathematical framework, namely with
matrices; in accelerator physics they are usually referred to as {\em transfer matrices.}
We only have to establish a few correspondences beforehand. For charged particles, 
the optical axis can be visualized as the center of the beam pipe. Moreover, at a 
longitudinal position $s,$ a particle can also be characterized by its distance 
$x$ and and angle $x'$ with respect to the center of the beam pipe. Here, we need
to point out that for charged particles this description is valid for moderately 
small angles $x',$ which is called the {\em paraxial approximation} and is valid 
for angles in the range of a few tens of mrad. But this is usually valid in
accelerators---10\,mrad correspond to a change of distance of 10\,mm per meter!
\par
In the absence of external influences---this can be visualized as a piece of beam 
pipe without magnets---charged particles also move on a straight line. The matrix 
that maps position $x$ and angle $x'$ from one end of the empty pipe to the other 
end is the same we used for light rays and is given in Equation~\ref{eq:drift}. 
This matrix is commonly referred to as the {\em matrix for a drift space} and the 
empty piece of beam pipe is referred to as a {\em drift space.} 
\par
The elements corresponding to the optical lenses are magnetic elements with a 
magnetic field that depends linearly on the transverse position $x.$ Note the
resemblance with the light rays: all parallel rays have to go through the focal
point downstream and that requires that the deflection angle of each ray has to
be proportional to its distance from the optical axis. In the same way must
charged particles be deflected more, due to the Lorentz force that the moving 
particles experience, if they are further from the axis of the magnet. Magnets with 
the required linearly rising magnetic field are {\em quadrupoles} and, if they are
moderately short, they can be described by the matrix in Equation~\ref{eq:quad}.
Their inverse focal length $1/f$ is directly proportional to the magnetic gradient 
$\pt B_y/\pt x,$ but we leave the details of the conversion to more advanced 
texts~\cite{VZAPB,WOLSKI,WIEDEMANN,SYLEE}.
\par
We note, however, that the magnetic fields in the quadrupoles have to fulfill
Maxwell's equations. This forces the transverse components $B_x$ and $B_y$ of 
the magnetic field to obey the curl equation: $\pt B_y/\pt x - \pt B_x/\pt y =0.$ 
A linearly increasing field component in one direction causes the other component 
to decrease. Quadrupoles therefore focus particles in one transverse plane, but 
defocus the same particles in the other plane. Since we will only consider one 
plane in this tutorial, we will not encounter this ``feature'' directly. We will, 
however, build beam lines consisting of both focusing and defocusing quadrupoles. 
The motivation to use both types of quadrupoles is to be able to describe the other
plane in the beam pipe as well.
\par
There are many more magnets that are used to guide the beam---big dipoles and small
steering magnets---and other magnets to correct various effects, but we will, in
the early sections of this tutorial, confine ourselves to drift spaces and quadrupoles 
and will build larger beam optical systems with these elements alone. We will 
implement most matrix manipulations in software in order to avoid excessive matrix 
calculations by hand.
\section{Software to move particles around}
Our task is to follow a particle through a sequence of drift spaces and quadrupoles,
modeled by the matrices from Equation~\ref{eq:drift} and~\ref{eq:quad}. To do so,
we first encode the matrices in a convenient way. Here we use octave's ability to 
write functions with the help of the \verb+@+-operator, which specifies the input
argument
\begin{verbatim}
  D=@(L)[1, L; 0, 1];
\end{verbatim} 
This command defines a function {\tt D()} that receives the length {\tt L} as input and 
returns the matrix from Equation~\ref{eq:drift}. We can then write {\tt D(2)*D(5)} at the 
octave command prompt and should obtain a $2\times 2$ matrix with {\tt 7} in the upper 
right corner. Likewise, we define the function {\tt Q()}
\begin{verbatim}
  Q=@(F)[1, 0; -1/F, 1]; 
\end{verbatim} 
which returns the matrix from Equation~\ref{eq:quad}. We also note that the column
vector with $\vec x_0=(x,x')$ can be expressed as
\begin{verbatim}
  X0=[x; xp];
\end{verbatim}
where the semicolon separates rows in a vector or matrix, and a comma separates the 
columns. The semicolon at the end of a command suppresses displaying the result of the
assignment or the calculation. 
\begin{itemize}
\item{\bf Exercise 5:} Set {\tt F=3} and numerically verify what you found in Exercise~4, 
  namely that parallel rays cross the axis after a distance $L=f.$
\end{itemize}
Using these matrices it is straightforward to find out what the transverse coordinates 
$(x,x')$ of a ray that enters the system
at one end will be, once it reaches the other end. To make this analysis more efficient 
we need a language to describe beam lines with many elements.
\par
We encode the description of the beam line in a matrix, whose first column contains
a code: we use {\tt 1} for a drift space and {\tt 2} for a quadrupole. The second column
contains a repeat code, which is mostly for cosmetic reasons in order to make plots look smoother.
A number 10 simply instructs the software to repeatedly apply the matrix for that element 
ten times. The third column contains the length of the element and the fourth column
contains an additional parameter; for a quadrupole, for example, it will contain its
focal length. An example of such a beam-line description is the following
\begin{verbatim}
  F=2.5;
  fodo=[1,  5,  0.2,  0;
        2,  1,    0,  F;
        1, 10,  0.2,  0; 
        2,  1,    0, -F;
        1,  5,  0.2,  0]; 
\end{verbatim}
which first defines the focal length {\tt F} to be 2.5\,m and then defines
an array, called {\tt fodo} that follows the conventions defined above. 
The first line describes 5 segments of a drift space, each being 0.2\,m long. The second 
line describes a quadrupole that is not segmented and therefore only has the repeat-count 
of~1. Moreover, the quadrupole is thin and has the length zero in the third column and the
focal length in the fourth. The third line describes a drift space, consisting of ten 
segments with length 0.2\,m each. The fourth line, again with code 2 in the first column,
describes a second quadrupole, this time with a negative focal length. The last line is
equal to the first one. We call the array that describes our beam line {\tt fodo}, because 
it consists of an alternating sequence of focusing (F) and defocusing (D) quadrupoles,
separated by not-focusing (O) drift spaces. FODO-based beam lines are very common, for 
example, the arcs of the LHC are based on them. The reason to alternate the quadrupole
polarity is to focus in both transverse planes equally, despite the each quadrupole treating
the two planes in the opposite manner.
\begin{itemize}
\item{\bf Exercise 6:} Recall that the imaging equation for a lens is $1/b+1/g=1/f,$
  which corresponds to a system of one focusing lens with focal length $f,$ sandwiched
  between drift spaces with length $g$ and $b,$ respectively. Write a beam-line
  description that corresponds to this system. We will later return to it and analyze it. 
\end{itemize}
Having a description of sequence of elements in a beam line available, we now will turn to
assembling all matrices that map the coordinates of a particles at the start of the beam 
line to each and every beam-line element. This will allow us to follow a particle on
its journey through the accelerator. 
\par
First we have to take care of some accounting and use the following code to accomplish that.
\begin{verbatim}
  beamline=fodo;               
  nlines=size(beamline,1);     
  nmat=sum(beamline(:,2))+1;   
  Racc=zeros(2,2,nmat);        
  Racc(:,:,1)=eye(2);          
  spos=zeros(nmat,1);          
\end{verbatim}
We use the convention to call the beam line
{\tt beamline} in the software. In this way we can simply redefine {\tt beamline} if we 
want to analyze a different one. Then we also need the number of lines {\tt nlines} in 
the beam-line description. We just use the {\tt size} of the first dimension of {\tt beamline},
which is the number of rows; for {\tt fodo} this would be 5. We also need the number of 
matrices {\tt nmat} we will have to calculate, which depends on the repeat-count in the 
second column of {\tt beamline}. We simply add these number and add one to place a unit
matrix in the first location, which makes array handling more convenient. Then we 
allocate space for all the matrices {\tt Racc}; we create an array of {\tt nmat}
$2\times 2$ matrices filled with zeros. Note that each $2\times 2$ matrix will eventually
contain the transfer matrix from the {\em start of the beam line} up to the location
immediately after {\em each segment.} We fill the first $2\times 2$ matrix with a unit
matrix, which is denoted by {\tt eye(2)} in octave or MATLAB. Finally, we allocate the 
array {\tt spos} that will be filled with the longitudinal position of the end point of
each segment. This will allow us to make plots with the horizontal axis denoting the 
longitudinal position, rather than the sequence number of the segment.
\par
With the accounting out of the way, we are now ready to assemble the transfer matrices
with the following code
\begin{verbatim} 
  ic=1;                         % element counter
  for line=1:nlines             % loop over input elements
    for seg=1:beamline(line,2)  % loop over repeat-count 
      ic=ic+1;                  % next element          
      Rcurr=eye(2);             % matrix in next element
      switch beamline(line,1)  
      case 1   % drift
        Rcurr=D(beamline(line,3));
      case 2   % thin quadrupole
        Rcurr=Q(beamline(line,4));   
      otherwise
        disp('unsupported code')
      end		  
      Racc(:,:,ic)=Rcurr*Racc(:,:,ic-1);    % concatenate 
      spos(ic)=spos(ic-1)+beamline(line,3); % position of element   
    end
  end
\end{verbatim} 
We first initialize the element counter {\tt ic}, which helps us to keep track of 
all the segments in the beam line. Then we loop over the lines in the beam-line
description {\tt beamline} and then, in a second loop, over the repeat-count. The
first thing we do is incrementing the element counter {\tt ic} and initialize the
matrix for the current segment {\tt Rcurr} to the unit matrix {\tt eye(2)}. The
{\tt switch} statement branches, depending on the code number in the first column 
of the current line, which denotes the type of element. If the code is {\tt 1}, we 
copy the matrix for a drift space with length given by the third column 
{\tt beamline(line,3)} to {\tt Rcurr}. If the code is {\tt 2}, we copy the 
matrix for a quadrupole with its focal length specified in the fourth column
to {\tt Rcurr}. If the first column contains anything else, we write a short
note to the display. At this point, {\tt Rcurr} contains the appropriate matrix 
for the currently considered segment and we left-multiply it to the matrix
{\tt Racc(:,:,ic-1)} that ends just upstream of the current segment and we then store the
result in {\tt Racc(:,:,ic)}. This procedure will recursively assemble all matrices
in {\tt Racc}. Likewise, we add the length of the current segment to {\tt spos}, 
which is successively filled with the positions at the exit of each segment.
Once the loops have finished, {\tt Racc} will contain all matrices that map the
coordinates of a particle at he start of the beam line to the exit point of each 
segment and {\tt spos} tells us where along the beam line that point is. 
\par
Note that for future use, we encapsulate most of the functionality to allocate
memory for the transfer matrices and the actual calculation of the matrices in
a separate function {\tt calcmat()}, which receives the {\tt beamline} as
input and returns the matrices and some accounting information in the following
form:
\begin{verbatim}
  [Racc,spos,nmat,nlines]=calcmat(beamline)
\end{verbatim} 
The code for {\tt calcmat()}, which is very similar to {\tt beamoptics.m}, is 
reproduced in Appendix~\ref{sec:code}.
\par
Now we are ready to use this code. As example, we map the initial coordinates
{\tt x0}, specified as a column vector, along the beam line and plot the 
transverse position {\tt x} as a  function of the longitudinal position $s.$
\begin{verbatim}
  x0=[0.001;0];         % 1 mm offset at start
  data=zeros(1,nmat);   % allocate memory
  for k=1:nmat          
    x=Racc(:,:,k)*x0;
    data(k)=x(1);       % store the position
  end
  plot(spos,1e3*data,'k');
  xlabel('s [m]'); ylabel(' x [mm]');
\end{verbatim}
After {\tt x0} is defined, we allocate the array {\tt data} to contain the variables
we want to display. Then we loop over all the segments in the beam line, of which
there are {\tt nmat}. In the loop, we multiply the initial coordinate {\tt x0} with
transfer matrix to the exit of each segment, which is stored in {\tt Racc}. Since we
want to display the transverse position, we copy the position {\tt x(1)} to {\tt data},
which will automatically have the same array dimensions as {\tt spos}, such that we can 
{\tt plot} the transverse position as a function of {\tt spos}. Note that we multiply
the contents of {\tt data} by $10^3$ in order to convert it to mm, which is also
annotated in the axes labels.
\begin{itemize}
\item{\bf Exercise 7:} Prepare initial coordinates that describe a particle that is
  on the optical axis, but has an initial angle $x'$ and plot the position $x$ along 
  the beam line.
\item{\bf Exercise 8:} Plot the angle $x'$ along the beam line.
\end{itemize}
The code only shows the trajectory in a single FODO cell. If we want to display the
trajectory, for example, through five consecutive cells, we only have to replace the 
definition of the {\tt beamline} early in the file by
\begin{verbatim}
  beamline=[fodo;fodo;fodo;fodo;fodo];  
\end{verbatim}
where we must use the semicolon as a separator, because we need to stack the beam line
descriptions one after the other and thereby create more lines. Optionally, we can
replace the previous command by
\begin{verbatim}
  beamline=repmat(fodo,5,1);
\end{verbatim}
where we use the built-in command {\tt repmat}.
\begin{itemize}
\item{\bf Exercise 9:} Plot both the position $x$ and the angle $x'$ through five cells.
\item{\bf Exercise 10:} Plot the position $x$ through 100 cells, play with different 
  values of the focal length {\tt F} and explore whether you can make the oscillations 
  grow.
\item{\bf Exercise 11:} Use the beam line for the imaging system you prepared in 
  Exercise~6 and launch a particle with $x_0=0$ and an angle of $x'_0=1\,$mrad at 
  one end. Verify that this particle crosses the center of the beam pipe at the 
  exit of the beam line, provided that $b,g,$ and $f$ satisfy the imaging equation
  that is shown in Exercise~6. 
\end{itemize}
Up to now the software allows us to map single particles through a beam line. But
a beam consists of more than a few particles and we have to have a look at how
to describe such ensembles of particles.
\section{Many particles, the beam}
Each of the many particles in a beam has its individual coordinates $(x,x').$ Under 
most circumstances are the positions $x$ and angles $x'$ distributed according to a normal,
or Gaussian, distribution. We create the coordinates of $N$ particles as a $2\times N$
array {\tt beam} that we fill with normally distributed random numbers with the 
help of the built-in function {\tt randn()}. The following commands can be found in
the script {\tt propagate\_beam.m.}
\begin{verbatim}
  Npart=10000;
  beam=randn(2,Npart);
\end{verbatim}
These random numbers have a mean of zero and unit rms, such that we scale them by multiplying
the first coordinate by the rms beam size {\tt sigx} and the second coordinate by the rms 
angular divergence {\tt sigxp}.
\begin{verbatim}
  sigx=1; x0=0;      % 1 mm beam size and offset x0
  sigxp=0.5; xp0=1;  % 0.5 mrad angular divergence and initial angle xp0
  beam(1,:)=sigx*beam(1,:)+x0;
  beam(2,:)=sigxp*beam(2,:)+xp0;
\end{verbatim}\label{code:sigdef}
where we also introduce the initial offsets {\tt x0} and {\tt xp0}, which are then the 
respective means of the distribution. We can now verify that the distribution has the 
desired properties by calculating the  {\tt beam\_position} and the {\tt beam\_size} via
\begin{verbatim}
  beam_position=mean(beam(1,:);
  beam_size=std(beam(1,:))
\end{verbatim}
where {\tt beam(1,:)} refers to the first row of the matrix {\tt beam}, which 
contains the positions of all particles. The built-in functions {\tt mean()} and 
{\tt std()} return the average and rms of the distribution, respectively.
\begin{itemize}
\item{\bf Exercise 12:} Calculate the angular divergence of the {\tt beam}.
\end{itemize}
\begin{figure}[tb]
\begin{center}
\includegraphics[width=0.47\textwidth]{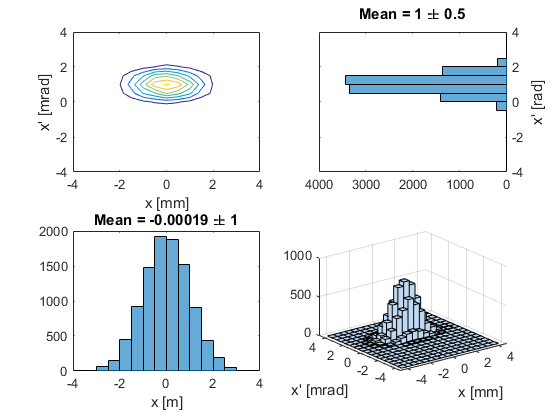}
\hskip 5mm
\includegraphics[width=0.47\textwidth]{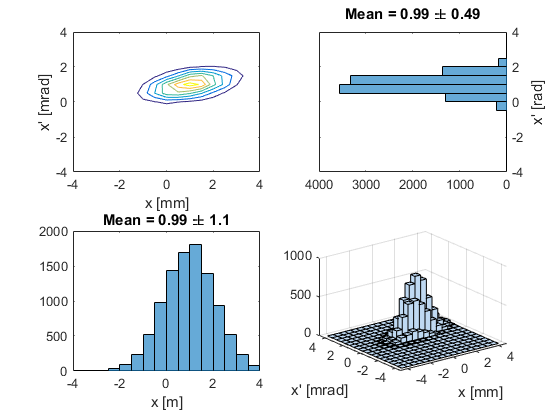}
\end{center}
\caption{\label{fig:beam}Particles at start (left) and after 1\,m drift space (right).}
\end{figure}
We can visualize the particle distribution by plotting one and two-dimensional 
histograms and contour plots. We encapsulate the code in the function {\tt show\_beam()}. 
The function, shown in its entirety in the appendix, receives the array {\tt beam} 
with the coordinates and the desired bin spacing for the histograms as input. It then 
creates four subplots: the top left shows a contour plot of the distribution in
$x$ and $x',$ the plots immediately to the right and below the contour plot show
the histograms with the projections onto the respective axes, and in the bottom right
a two-dimensional histogram is displayed.
\par
We then use a transfer matrix {\tt R} to propagate the {\tt beam} to a downstream
location, here through a drift space with a length of 1\,m, by multiplying the {\tt beam}
with {\tt R} , as shown in the following example.
\begin{verbatim}
  R=[1,1;0,1];  % drift space with L=1 m
  beam=R*beam;  % propagate particles
\end{verbatim}
Calling {\tt show\_beam()} again produces the plot on the right-hand side in 
Figure~\ref{fig:beam}. We find that the contour plot has slightly changed shape;
the ellipses have rotated. Moreover, the mean of the horizontal distribution has 
changed to 1\,mm. This is a consequence of the initial angle $x'_0=1\,$mrad, which
caused the beam to move towards the positive x-axis by 1\,mm over the 1\,m long 
drift space. The horizontal width has increased to 1.1\,mm.
\par
Note that we specified the transverse dimensions and angle in mm and mrad instead
of m and rad. This is permissible, because transfer matrices describe a {\tt linear}
transformation, and multiplying the vector to the right by some number, here $10^3,$
causes the output on the left of the matrix multiplication to be scaled by the same 
number.
\begin{itemize}
\item{\bf Exercise 13:} Try this out yourself:  Scale the input vector by 17 times 
  the month of your birthday (85 if you are born in May) and verify that the output
  vector from the matrix multiplication has changed by the same factor.
\end{itemize}
\par
\begin{figure}[tb]
\begin{center}
\includegraphics[width=0.8\textwidth]{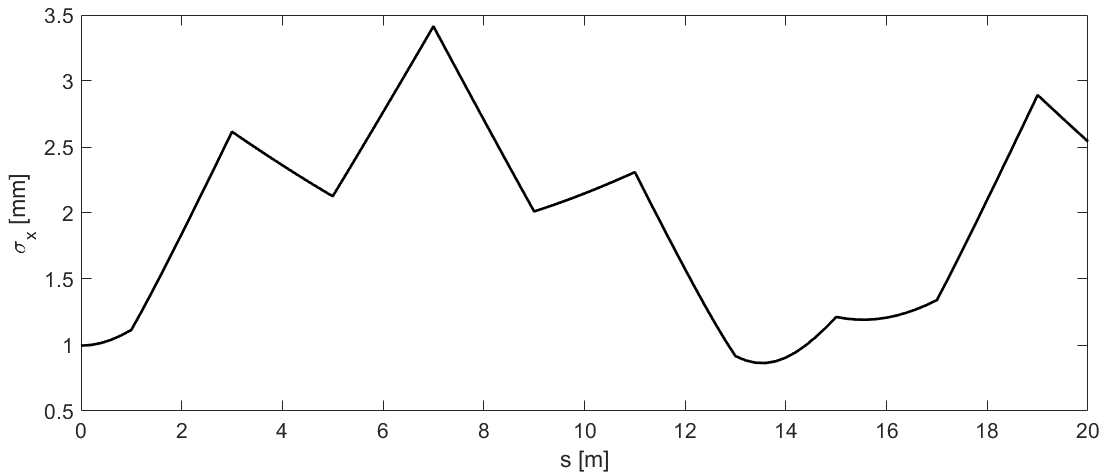}
\end{center}
\caption{\label{fig:sigvzs}The beam size along the beam line made of five FODO cells.}
\end{figure}
We will continue to use mm and mrad for the transverse coordinates and will now 
calculate the beam size after every segment in the beam line. We use {\tt beamoptics.m} 
to build a beam line of five consecutive FODO cells and prepare all transfer matrices
{\tt Racc}. Then we execute the following code:
\begin{verbatim}
  beam0=beam;           % copy initial distribution to beam0
  data=zeros(nmat,1);   % allocate space for plotting
  for k=1:nmat
    beam=Racc(:,:,k)*beam0;    % propagate beam0
    data(k,1)=std(beam(1,:));  % calculate rms beam size
  end
  plot(spos,data(:,1),'k')     % plot along the beam line
  xlabel('s [m]'); ylabel('\sigma_x [mm]')
\end{verbatim}\label{code:prop_a}
First we copy the beam distribution to {\tt beam0} in order to avoid overwriting it.
Then we allocate some space to store the values we intend to display and then loop
over all elements in the beam line. At each step  in the loop, we
propagate the initial {\tt beam0} to the new location in the beam line and store the 
rms of the positions of all particles in {\tt data}. After the loop completes, we 
display the beam size as a function of $s$ and annotate the axes. Figure~\ref{fig:sigvzs}
shows the resulting plot. We observe that the beam size grows over the first meter, where
it traverses a defocusing quadrupole, which increases the beam size further until at 
$s=3\,$m. There the first focusing quadrupole reduces the beam size until the defocusing quadrupole 
at $s=5\,$m increases the beam size again, whence the focusing quadrupole at $s=7\,$m 
reduces it again; and so forth. We also observe that the {\tt beam0} we initially
prepared can become larger up to an rms  beam size of a little under $\sigma_x=3\,$mm.
\begin{itemize}
\item{\bf Exercise 14:} Display (a) the average position of the particles along the 
  beam line. Likewise, display (b) the angular divergence.
\item{\bf Exercise 15:} What happens if you (a) reduce or (b) increase the initial
  angular divergence {\tt sigxp} by a factor of two? 
\end{itemize}
Normally, one is not really interested in the histogram of the particle distribution.
Information about the centroid position, the beam size, and possibly the angular 
divergence are enough. So, wouldn't it be nice, 
if we could move the interesting quantities around, instead of thousands of sample
particles? It turns out that this is possible and that is the topic of the next section.
%
%
\section{Moving beams around}
To simplify the writing of many matrix-valued equations, we introduce the notation 
that the position $x$ is denoted by $x_1$ and the angle $x'$ by $x_2.$ This allows us
to express the propagation of the particle coordinates as $\hat x_i=\sum_{j=1}^2 R_{ij}x_j.$
If we have to deal with many particles we label them by a second subscript, separated from 
the first by a comma; $x_{2,17}$ is thus the angle of particle number~17.
\par
When we calculated the average beam position using the built-in functions, the software 
actually performed the following operations
\begin{equation}
X_1=\langle x_1\rangle = \frac{1}{N}\sum_{m=1}^N x_{1,m}
\qquad\mathrm{and}\qquad
X_2=\langle x_2\rangle = \frac{1}{N}\sum_{m=1}^N x_{2,m}\ ,
\end{equation}
where the index $m$ sums over the $N$ particles. The notation with the angle brackets 
denotes thus averaging over the ensemble of particles. We also introduced the quantities
$X_1$ and $X_2$ to denote the averages. Likewise, the rms quantities are calculated via
\begin{eqnarray}
\sigma_x^2&=& \sigma_{11}=\langle(x-X_1)^2\rangle = \frac{1}{N}\sum_{m=1}^N (x_{1,m}-X_1)^2\nonumber\\
\sigma_{x'}^2&=& \sigma_{22}=\langle(x'-X_2)^2\rangle = \frac{1}{N}\sum_{m=1}^N (x_{2,m}-X_2)^2\ .
\end{eqnarray}
By inspection, we should not be surprised that there is also a third variant of the above sums,
given by
\begin{equation}
\sigma_{12}=\langle(x-X_1)(x'-X_2)\rangle = \frac{1}{N}\sum_{m=1}^N (x_{1,m}-X_1)(x_{2,m}-X_2)\ ,
\end{equation}
which describes the correlation between position (index 1) and angle (index 2). This quantity
$\sigma_{12}$ is actually capable of describing the rotation of the contour in the right-hand
plot in Figure~\ref{fig:sigvzs} that showed up after propagating the initial beam through
a 1\,m long drift space.
\par
These five quantities $X_1, X_2, \sigma_{11}, \sigma_{12},$ and $\sigma_{22}$ describe all 
the interesting properties of a beam. Note that $X_1$ and $X_2$ are the first moments of
the two-dimensional beam distribution and the $\sigma_{ij}$ with $i=1,2$ are the three 
independent second moments; actually they are the central moments, because they 
describe the second moments with respect to the centroids. Note that usually the
three independent second moments can be written to form a symmetric $2\times 2$--matrix
that is given by
\begin{equation}\label{eq:bmat}
\sigma=
\left(\begin{array}{cc} \sigma_{11} & \sigma_{12} \\ \sigma_{21}& \sigma_{22}\end{array}\right)\ ,
\end{equation}
with $\sigma_{21}=\sigma_{12}.$ We use the convention that sigmas without 
subscripts denote the matrix and sigmas with subscripts denote the matrix elements. 
We also point out that the sigma matrix changes from one location $s$ in the beam line 
to another location $\hat s.$ Remember that Figure~\ref{fig:sigvzs} shows the beam size
$\sigma_x=\sqrt{\sigma_{11}}$ as a function of $s$ along the beam line.
\par
In the more advanced literature~\cite{VZAPB,WOLSKI}, but also in Appendix~\ref{sec:propmom}, 
it is shown that the first and second moments propagate according to 
\begin{equation}\label{eq:prop}
\vec X(s_2) =  R \vec X(s_1)
\qquad\mathrm{and}\qquad
\sigma(s_2) = R \sigma(s_1) R^t\ ,
\end{equation}
where $\vec X(s)$ denotes the column vector with entries $X_1$ and $X_2$ at longitudinal
location $s.$ The first equation describes the remarkable fact that the centroid of
the beam---the first moments---propagate in the same way as single particles. The
second equation describes the propagation of the beam matrix, as defined in 
Equation~\ref{eq:bmat}. In Equation~\ref{eq:prop} we use the common convention to
denote the transpose of the matrix $R$ by $R^t.$
\par
\begin{figure}[tb]
\begin{center}
\includegraphics[width=0.8\textwidth]{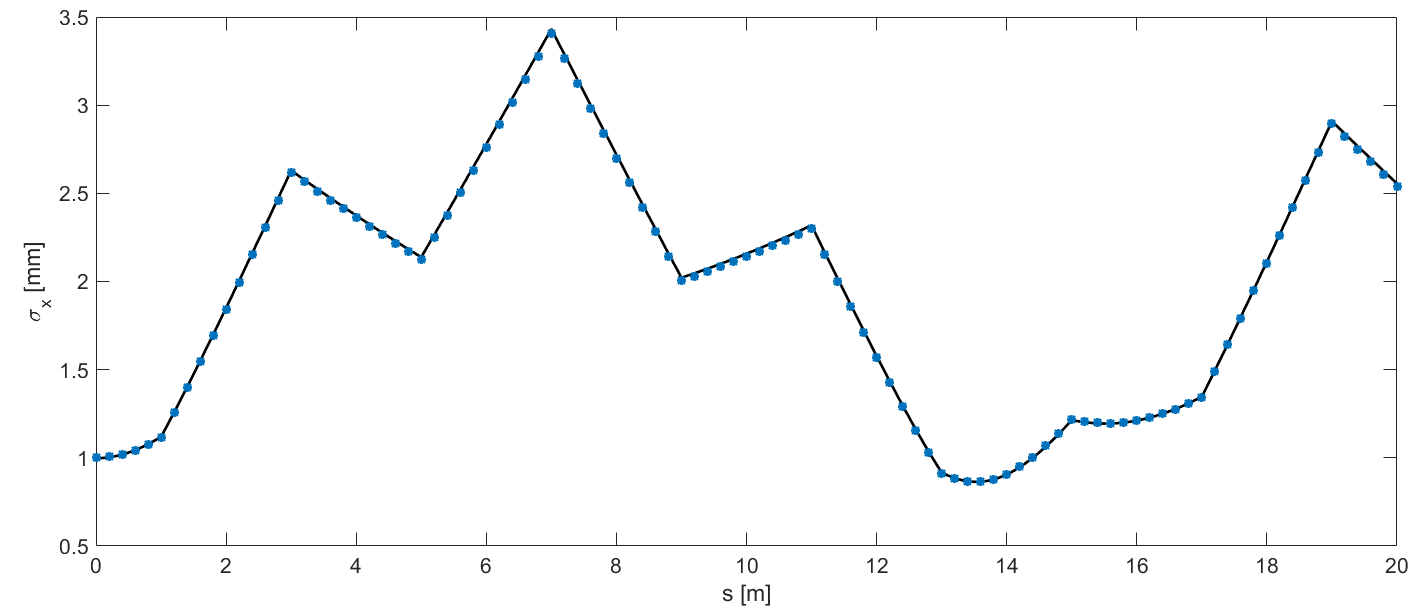}
\end{center}
\caption{\label{fig:sig2}The beam size (line) from Figure~\ref{fig:sigvzs} and calculated
  from Equation~\ref{eq:prop} (asterisks).}
\end{figure}
The two equations in Equation~\ref{eq:prop} are particularly convenient and efficient to 
implement in software, because they just describe matrix multiplication of the centroid 
$\vec X$ and the sigma matrix $\sigma$, both of which describe the beam, and the transfer 
matrices $R,$ which describe the hardware of the beam line. We can now implement these
equations in the software by adding the following lines after the code from page~\pageref{code:prop_a},
which is re-used to define the initial beam and assemble the transfer matrices {\tt Racc}.
The following code segment then adds the calculations according to Equation~\ref{eq:prop}.
\begin{verbatim}
  X0=[0;xp0];                      % 1 mrad initial angle 
  sigma0=[sigx^2,0;0,sigxp^2];   % initial sigma matrix
  for k=1:nmat
    X=Racc(:,:,k)*X0;                      % Equation 7
    sigma=Racc(:,:,k)*sigma0*Racc(:,:,k)';
    data(k,2)=sqrt(sigma(1,1));
  end
  plot(spos,data(:,1),'k',spos,data(:,2),'*')
  xlabel('s [m]'); ylabel('\sigma_x [mm]')
\end{verbatim}
In this code segment, we first initialize the centroid {\tt X0} and the sigma-matrix 
{\tt sigma0} with the same values used on page~\pageref{code:sigdef}. Then we loop
over the {\tt nmat} positions in the beam line and propagate both centroid and
sigma-matrix according to Equation~\ref{eq:prop}. In this example we only store
the beam size $\sigma_x=\sqrt{\sigma_{11}}$ in the second column of {\tt data}.
After the loop completes, we plot both the newly calculated beam sizes and those
calculated previously as a function of {\tt spos}. The result, displayed in 
Figure~\ref{fig:sig2}, shows very good agreement of the beam sizes. Henceforth, we
will only use Equation~\ref{eq:prop} for propagating beams through beam lines.
\begin{itemize}
\item{\bf Exercise 16:} Using Equation~\ref{eq:prop}, display (a) the average position 
  of the particles along the beam line. Likewise, (b) display the angular divergence.
  Compare with the result you found in Exercise~14.
\item{\bf Exercise 17:} Can you find an initial beam matrix {\tt sigma0}
  that reproduces itself at the end of the beam line?
\end{itemize}
In Figure~\ref{fig:sig2} the beam size $\sigma_x$ oscillates in a somewhat uncontrolled 
fashion along the beam line. Next we will explore whether we can find periodic oscillations
that repeat after each cell. 
\section{Periodic systems and beams}
To explore the periodicity of a beam optical system, we first consider a single FODO
cell only and follow a single particle with initial coordinates {\tt x=[1;0]} over 
a large number of turns, as shown in the following script
\begin{verbatim}
  % trak_single_particle.m
  beamoptics;       %  with beamline=fodo;
  Nturn=100; data=zeros(Nturn,2);
  x=[1;0];    % initial condition, 1 mm, no angle
  for k=1:Nturn
    x=Racc(:,:,end)*x;
    data(k,1)=x(1); data(k,2)=x(2);
  end
  plot(data(:,1),data(:,2),'.')
  xlabel( 'x [mm]'); ylabel('x'' [mrad]')
\end{verbatim}
\begin{figure}[tb]
\begin{center}
\includegraphics[width=0.6\textwidth]{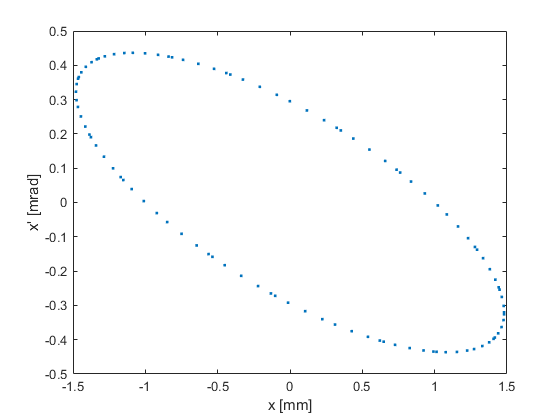}
\end{center}
\caption{\label{fig:ps} A phase-space plot showing positions $x$ versus angles $x'$ of 
  a particle followed for 100 consecutive turns.}
\end{figure}
Before running this script, we have to update {\tt beamoptics.m} to prepare matrices
for a single cell by defining {\tt beamline=fodo}, such that calling {\tt beamoptics}
makes all matrices available to the script shown on this page. After specifying the 
requested number of 
turns {\tt Nturn} and allocating the array {\tt data}, we start the loop. Inside the
loop, we use the present coordinates {\tt x}, map them through one turn by multiplying
with the transfer matrix {\tt Racc(:,:,end)}, and store the coordinates in the same
vector {\tt x}, which is thereby continuously updated with the coordinates turn after 
turn. In each turn, we also save the coordinates to {\tt data} for later display.
After the loop has finished, we plot the first column of {\tt data}, which contains 
the positions $x$ on consecutive turns, versus the second column, which contains
the angle $x'$ on the corresponding turns. The plot, shown in Figure~\ref{fig:ps},
displays an ellipse. 
\begin{itemize}
\item{\bf Exercise 18:} Explore different initial coordinate and compare the phase-space
  plots you obtain.
\item{\bf Exercise 19:} Execute {\tt plot(data(:,1))} to observe the turn-by-turn positions.
  What do you observe? 
\item{\bf Exercise 20:} In the definition of {\tt fodo} at the top of {\tt beamoptics.m}
  reverse the polarity of both quadrupoles and prepare a phase-space plot. How does it
  differ from the one in Exercise~18?
\item{\bf Exercise 21:} Prepare an array describing a FODO cell that starts immediately
  following the quadrupole with the negative focal length and prepare the phase-space
  plot. 
\end{itemize}
The particle, performing stable oscillations reminiscent of a harmonic oscillator, 
suggests that we can split the transfer matrix for one turn into a product of three 
matrices~\cite{VZAPB}; a rotation matrix, sandwiched between two matrices that distort 
the coordinates
\begin{equation}\label{eq:RAOR}
R={\cal A}^{-1}{\cal O}{\cal A}
\quad\mathrm{with}\quad
{\cal O}=\left(\begin{array}{rr} \cos\mu &\sin\mu\\ -\sin\mu & \cos\mu\end{array}\right)
\quad\mathrm{and}\quad
{\cal A}=
\left(\begin{array}{cc} \frac{1}{\sqrt{\beta}} & 0 \\ 
\frac{\alpha}{\sqrt{\beta}} & \sqrt{\beta}\end{array}\right)\ .
\end{equation}
Here $\alpha$ and $\beta$ are two, presently undetermined, parameters that describe the 
distortion. We can, however, determine them from $R$ by comparing coefficients with
the result 
\begin{equation}\label{eq:betaR}
\mu=\arccos\left(\frac{R_{11}+R_{22}}{2}\right)\ ,
\quad \beta=\frac{R_{12}}{\sin\mu}\ ,
\quad\mathrm{and}\quad \alpha=\frac{R_{11}-R_{22}}{2\sin\mu}\ .
\end{equation} 
These equations are encapsulated in the following function {\tt R2beta()}
\begin{verbatim}
  function [Q,alpha,beta,gamma]=R2beta(R)
  mu=acos(0.5*(R(1,1)+R(2,2)));
  if (R(1,2)<0) mu=2*pi-mu; end
  Q=mu/(2*pi);
  beta=R(1,2)/sin(mu);
  alpha=(0.5*(R(1,1)-R(2,2)))/sin(mu);
  gamma=(1+alpha^2)/beta;
\end{verbatim}
which allows us to determine the parameters $\alpha, \beta,$ and $\mu=2\pi Q$ from any
transfer matrix $R$ that permits stable oscillations. Note that this is only possible
for $\vert (R_{11}+R_{22})/2\vert <1,$ which indicates the limit of stability.
\begin{itemize}
\item{\bf Exercise 22:} Find the range of focal lengths {\tt F} for which the FODO
  cells permit stable oscillations.
\end{itemize}
The parameters returned by {\tt R2beta()} are commonly called the phase advance $\mu,$
the Twiss parameters $\alpha,\beta,$ and $\gamma=(1+\alpha^2)/\beta,$ and $Q$ is called the tune.
\par
So, why did we analyze the transfer matrix if we actually want to construct a 
sigma matrix that is periodic? Because the Twiss parameters make it particularly
simple to build a beam matrix {\tt sigma0} that is repeats itself after one cell.
It is given by
\begin{equation}\label{eq:sigma0}
\sigma_0=\eps \left(\begin{array}{rr} \beta & -\alpha \\ -\alpha & \gamma \end{array}\right)\ ,
\end{equation} 
where we introduced the emittance $\eps,$ whose relevance will become obvious in a short
while. We can now propagate the matrix $\sigma_0$ through the cell and use Equation~\ref{eq:RAOR} 
to express the transfer matrix $R$ through ${\cal A}$ and ${\cal O}.$ After some algebra, we
find
\begin{equation}
R\sigma_0R^t = \sigma_0\ ,
\end{equation}
which shows that the beam matrix, as defined in Equation~\ref{eq:sigma0}, reproduces itself 
after one cell. Let's try out whether this really works in the following code.
\begin{verbatim}
  % periodic_beam.m
  beamoptics;   % with beamline=fodo;
  [Q,alpha,beta,gamma]=R2beta(Racc(:,:,end));
  data=zeros(nmat,1);  % allocate memory for display
  eps=1; % set emittance to unity
  sigma0=eps*[beta, -alpha;-alpha,gamma];
  for k=1:nmat
    sigma=Racc(:,:,k)*sigma0*Racc(:,:,k)';
    data(k,1)=sqrt(sigma(1,1));
  end
  plot(spos,data(:,1),'k') 
\end{verbatim}
Here we call {\tt beamoptics} to prepare the transfer matrices and then {\tt R2beta()} 
to calculate the Twiss parameters, from which we build the initial beam matrix {\tt sigma0}
that is subsequently propagated through the beam line.
\begin{itemize}
\item{\bf Exercise 23:} Run {\tt periodic\_beam} and convince yourself that the sigma matrix
  at the end of the cell is indeed equal to {\tt sigma0}.
\item{\bf Exercise 24:} Write down the numerical values of initial beam matrix {\tt sigma0}, then
  build a beam line made of 15 consecutive cells by changing the definition of {\tt beamline}
  in {\tt beamoptics.m} and then, using {\tt sigma0} with the noted-down numbers, prepare
  a plot of the beam sizes along the 15 cells. Is it also periodic?
\end{itemize}
The emittance $\eps$ that appeared in Equation~\ref{eq:sigma0} as an arbitrary constant is
actually conserved when propagating the beam matrix, which is easily seen from 
\begin{equation}
\det(\sigma) = \det(\tilde R\sigma_0\tilde R^t) =
\det(\tilde R)\det(\sigma_0)\det(\tilde R^t) = \det(\sigma_0)=\eps^2
\end{equation}
for an arbitrary transfer matrix $\tilde R,$ because all transfer matrices have unit 
determinant. This is true for all maps that describe the motion of a beam through static
magnetic fields, in particular, for the matrices for drift space and quadrupole and 
therefore also for products of those elements. 
\begin{itemize}
\item{\bf Exercise 25:} Verify that all matrices {\tt Racc(:,:,k)} have unit determinant.
\end{itemize}
Note also that the matrix with the Twiss parameters $\alpha,\beta,$ and $\gamma,$ which
appears in Equation~\ref{eq:sigma0}, has
unit determinant by construction. The Twiss parameters $\alpha, \beta,$ and $\gamma$ 
describe the relative magnitude of the matrix elements in $\sigma_0$ and the
emittance $\eps$ determines the absolute magnitude of the beam. In particular, when
considering the 11-element of a sigma matrix, we see that the absolute beam size is
given by $\sigma_x^2=\sigma_{11}=\eps\beta.$ Here the emittance $\eps$ is constant 
and sets the scale, whereas $\beta(s)$ describes the modulation of the beam size
along a beam line from one position $s$ to another.
\begin{itemize}
\item{\bf Exercise 26:} Multiply {\tt sigma0} from Exercise~24 by 17 and calculate the
  emittance. Then propagate the sigma matrix through the beam line from Exercise~24 and
  verify that the emittance of the sigma matrix after every element is indeed constant
  and equal to its initial value.
\end{itemize}
Up to now, we have used FODO cells and calculated the phase advance $\mu$ and the 
Twiss parameters from the transfer matrices. We now reverse the procedure and try
to find hardware parameters, such as the focal length of a quadrupole, to achieve
certain, desirable values. This procedure, commonly called {\em matching,} is the
topic of the next section.
\section{Satisfying one's desires: matching}
From the analysis of large circular accelerators it is known that an important parameter 
for the stability of periodically traversed beam lines is the phase advance $\mu$ of
a cell, or the tune $Q,$ which is the phase advance for the whole ring, 
divided by $2\pi.$ Let us consider a single cell first and try to adjust the
focal length {\tt F} of the quadrupoles to a value, such that the phase advance of
the cell is $\mu=60^o,$ which corresponds to $Q=\mu/2\pi=1/6.$ The following
short script first defines {\tt F} and {\tt fodo,} then calls {\tt calcmat()} to
calculate the transfer matrices, and finally uses the last one {\tt Racc(:,:,end)}
as argument for {\tt R2beta()}, which returns the phase advance and the Twiss 
parameters. 
\begin{verbatim}
  % match_phase_advance.m
  clear all; close all
  F=2.5;   % focal length of the quadrupoles
  fodo=[ 1,  5,  0.2,  0; 
         2,  1,  0.0, -F; 
         1, 10,  0.2,  0; 
         2,  1,  0.0,  F; 
         1,  5,  0.2,  0];
  beamline=fodo;
  [Racc,spos,nmat,nlines]=calcmat(beamline);
  [Q,alpha,beta,gamma]=R2beta(Racc(:,:,end));
  Q=Q
\end{verbatim}
The statement {\tt Q=Q} serves only to display the value of {\tt Q} which is 
the sought-after phase advance $\mu$ divided by $2\pi.$
\begin{itemize}
\item{\bf Exercise 27:} Vary {\tt F} by hand and try to (a) find a value that 
  returns $Q=1/6.$ (b) Then try to find a value of {\tt F} that produces a $90^0$
  phase-advance. What is the corresponding value of $Q$?
\end{itemize}
This procedure mimics the matching done by the accelerator-design software. There
a minimizer optimizes a cost-function that encodes the desired constraints. In
Exercise~27 you have to do it by hand.
\par
\begin{figure}[tb]
\begin{center}
\includegraphics[width=0.8\textwidth]{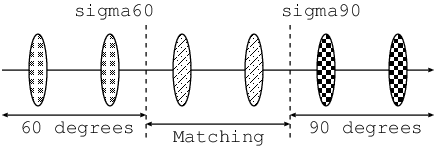}
\end{center}
\caption{\label{fig:matching}Illustration for Exercise 28, where the strength of 
  the two matching quadrupoles needs to be determined.}
\end{figure}

A common task when designing accelerators is matching one section of a beam line
to another one. Here we will assume that the upstream beam line consists of
FODO cells with a $60^o$ phase advance and the downstream beam line of FODO cells 
with a $90^o$ phase advance. These are the cells with the focal length we calculated
in Exercise~27. In between the $60^o$ and $90^o,$ we place a third cell with two
quadrupoles that we will use to match the upstream to the downstream beam line.
To do so, we need to prepare periodic beam matrices {\tt sigma60} and
{\tt sigma90} for the respective sections. Note that {\tt sigma90} only depends 
on {\em two} parameters: the Twiss parameters $\alpha$ and $\beta,$ and therefore we 
also need {\em two} quadrupoles with independently variable focal length to adjust
until the final beam matrix equals {\tt sigma90}. See Figure~\ref{fig:matching}
for an illustration of the geometry.
\begin{itemize}
\item{\bf Exercise 28:} Implement the procedure described in the previous paragraph.
\end{itemize}
The discussed matching tasks are only basic examples of the tasks encountered when
designing an accelerator. As mentioned before, most software packages have a 
more or less convenient way to specify the constraints to fulfill and the magnets
to vary to satisfy the constraints. Then the software automatically produces the
magnet settings. 
%
%
\section{...and beyond...}
So far we only considered the very basic optical elements and confined ourselves to
a single transverse dimension. During CAS we will build on these basics and extend
the software in several directions. In this section, we therefore give some
indication of things to come. We start by including quadrupoles that have a finite
length. In the introductory beam optics lectures you will learn that the matrix for
these elements depends on the polarity of the magnet excitation, as specified by a
parameter $k_1\propto \pt B_y/\pt x.$ The matrix is given by
\begin{equation}\label{eq:QF}
Q(l,k_1)=\left(\begin{array}{cc}
\cos(\sqrt{k_1}l) & \frac{1}{\sqrt{k_1}}\sin(\sqrt{k_1}l) \\
-\sqrt{k_1}\sin(\sqrt{k_1}l) & \cos(\sqrt{k_1}l)\\
\end{array}\right)
\quad\mathrm{for}\quad k_1 \geq 0
\end{equation} 
and 
\begin{equation}\label{eq:QD}
Q(l,k_1)=\left(\begin{array}{cc}
\cosh(\sqrt{\vert k_1\vert}l) & \frac{1}{\sqrt{\vert k_1\vert}}\sinh(\sqrt{\vert k_1\vert}l) \\
\sqrt{\vert k_1\vert}\sinh(\sqrt{\vert k_1\vert}l) & \cosh(\sqrt{\vert k_1\vert}l)\\
\end{array}\right)
\quad\mathrm{for}\quad k_1 < 0
\end{equation}
and this brings us directly to 
\begin{itemize}
\item{\bf Exercise 29:} Include these matrices in {\tt calcmat()} and assign code {\tt 5} to 
  them. Hint: write an external function that returns $Q(l,k_1)$ and call it from {\tt calcmat()}. 
\end{itemize}
Once the software is able to handle long quadrupoles, we can replace the thin quadrupoles
in the beam-line description used earlier.
\begin{itemize}
\item{\bf Exercise 30:} Use the beam line from Exercise~27 (60 degrees/cell FODO) and 
  replace the thin quadrupoles by long quadrupoles with a length of 0.2, 0.4, 1.0\,m. 
  Make sure the overall length and the phase advance of the FODO cell remains unchanged.
  By how much does the periodic beta function at the start of the cell change? Express 
  the change in percent. 
\end{itemize}
In the introductory optics lectures you will learn that a so-called sector dipole magnet
will exhibit {\em weak focusing} in the bending plane. The transfer matrix for this 
magnet will be shown to be 
\begin{equation}\label{eq:BB}
B(l,\phi)=\left(\begin{array}{cc}
\cos(\phi) & \rho\sin(\phi) \\ -\sin(\phi)/\rho & \cos(\phi)
\end{array}\right)
\quad\mathrm{with}\quad \rho=l/\phi
\end{equation}
\begin{itemize}
\item{\bf Exercise 31:} Include this matrix in {\tt calcmat()} and assign the code {\tt 4} 
  to it.
\item{\bf Exercise 32:} Insert 1\,m long dipoles in the center of the drift spaces of the 
  FODO cells from Exercise~27 while keeping the length of the cell constant. Investigate 
  deflection angles of $\phi=5,10,$ and $20\,$degrees. 
  Check by how much the periodic beta functions change. Why do they change? Explain! Can 
  you compensate the phase advance $\mu$ by adjusting the strength or focal lengths of 
  the quadrupoles?
\end{itemize}
All matrices discussed up to now only address one transverse plane and assume that all
particles have the same momentum $p_0$. 
\par
But this is only approximately true in a real 
accelerator and we therefore must take small momentum deviations $\delta = (p-p_0)/p_0$
into account. Instead of describing a particle by its transverse coordinates $(x,x')$
alone, we now characterize each particle by $(x,x',\delta),$ which requires to upgrade
the transfer matrices to $3\times 3$ matrices that operate on state vectors $(x,x',\delta).$
The matrix for a drift space therefore turns out to be
\begin{equation}
D(L)=\left(\begin{array}{ccc} 1 &L & 0\\ 0 & 1 & 0 \\ 0 & 0 & 1\end{array}\right)
\end{equation} 
with the original $2\times2$ matrix from Equation~\ref{eq:drift} in he top left corner,
a $1$ in the bottom right corner and zeros in the other positions of the third row and
the third column. The transfer matrices for thin and long quadrupoles are updated likewise.
Only the matrix for the sector dipole has non-zero entries in the third column
\begin{equation}
B(l,\phi)=
\left(\begin{array}{ccc}
\cos\phi & \rho\sin\phi      & \rho(1-\cos\phi) \\
-\sin(\phi)/\rho & \cos\phi  & \sin\phi \\
        0        &    0      & 1  \\
\end{array}\right)
\end{equation}
In order not to mix functions with the same name, but using $3\times3$ instead of 
$2\times 2$ matrices, prepare a subdirectory for the upgraded matrices.
\begin{itemize}
\item{\bf Exercise 33:} Upgrade the software to consistently handle $3\times3$ 
  matrices for drift space, quadrupoles, and sector dipoles.
\item{\bf Exercise 34:} Build a beam line of six FODO cells with a phase advance of 
  60 degrees/cell (thin quadrupoles are OK to use) and add a sector bending magnet 
  with length 1 m and bending angle $\phi=10\,$degrees in the center of each drift. 
  You may have to play with the quadrupole values to make the phase advance close 
  to 60\,degrees. But you probably already did this in Exercise~32.
\item{\bf Exercise 35:} Use the starting conditions $(x_0,x_0',\delta)=(0,0,0)$ and 
  plot the position along the beam line. Repeat this for $\delta=10^{-3}$ and for 
  $\delta=3\times 10^{-3}$. Plot all three traces in the same graph. Discuss what 
  you observe and explain!
\end{itemize}
We also need to upgrade the beam matrix $\sigma$ to be consistent with the 
$3\times 3$ transfer matrices. It is 
then a $3\times 3$ matrix itself with the original $2\times 2$ sigma-matrix from 
Equation~\ref{eq:bmat} in the top-left corner, zeros in the third column and row,
except the matrix element in the lower-right corner, which contains the square of
the momentum spread $\sigma_p.$ 
\begin{itemize}
\item{\bf Exercise 36:} Work out the transverse components of the periodic beam
  matrix $\sigma_0$. Assume that the emittance is $\eps_0=10^{-6}$\,meter-rad. 
  Furthermore, assume that the momentum spread $\sigma_0(3,3)=\sigma_p^2$ is zero 
  and plot the beam size along the beam line.
\item{\bf Exercise 37:} Plot the beam size for for $\sigma_p^2=10^{-3}$ and for  
  $\sigma_p^2=3\times 10^{-3}$. What happens if you change the phase advance of 
  the cell? Try out by slightly changing the focal lengths.
\item{\bf Exercise 38:} Determine the periodic dispersion at the start of the 
  cell. Then plot the dispersion in the cell. 
\end{itemize}
For these exercises you definitely need to consult the slides from the introductory
beam optics lectures for background information.
\par
Instead of adding the momentum deviation as additional degree of freedom for the 
simulation, we can extend the simulation to comprise of both transverse degrees
of freedom, the horizontal and the vertical plane. The matrices that describe
the dynamics are then $4\times4$ matrices with the matrices for the respective
planes filling the upper-left and the lower-right $2\times 2$ sub-matrices. As 
an illustration, we show the matrix for a focusing ($k_1>0$) quadrupole
\begin{equation}
Q(l,k_1)=\left(\begin{array}{cc} Q_F & 0_2 \\ 0_2 & Q_D \end{array}\right)\ ,
\end{equation} 
where $Q_F$ is the matrix from Equation~\ref{eq:QF}, $Q_D$ from Equation~\ref{eq:QD},
and $0_2$ is the $2\times2$ matrix containing only zeros. For $k_1<0$ the sub-matrices
$Q_F$ and $Q_D$ are exchanged. The matrix for a thin quadrupole is constructed likewise.
From the introductory lectures, we know that the matrix for sector dipole only shows
weak focusing in the horizontal plane, but behaves as a drift space in the vertical 
plane. Its matrix is therefore given by
\begin{equation}
B(l,\phi)= \left(\begin{array}{cc} B_2 & 0_2 \\ 0_2 & D_2 \end{array}\right)\ ,
\end{equation}
where $B_2$ is the $2\times 2$ matrix from Equation~\ref{eq:BB} and $D_2$ the matrix
for  a drift space from Equation~\ref{eq:drift}. Note that in this framework, we
assume that all particles have the reference momentum $p_0$ and we can therefore
ignore the momentum deviation~$\delta.$
\begin{itemize}
\item{\bf Exercise 39:} Convert the code to use $4\times4$ matrices, where the 
  third and fourth columns are associated with the vertical plane. Create a separate 
  subdirectory for the calculations with the $4\times 4$ matrices. 
\end{itemize}
Using this software to address the following exercises. Note that changing a quadrupole 
simultaneously affects the phases advances $\mu_x, \mu_y$ and the beta functions
in the respective planes.
\begin{itemize}
\item{\bf Exercise 40:} Start from a single FODO cell with 60 degrees/cell you used 
  earlier. Insert sector bending magnets with a bending angle of $\phi=10\,$degrees in the 
  center of the drift spaces. The bending magnets will spoil the phase advance in 
  one plane. Now you have two phase advances and need to adjust both quadrupoles 
  (by hand to 2 significant figures) such that it really is 60 degrees in both planes. 
\item{\bf Exercise 41:} Use the result from exercise 2 and adjust the two quadrupoles 
  such that the phase advance in the horizontal plane is 90 degrees, cell, while it 
  remains 60 degrees/cell in the vertical plane. 
\item{\bf Exercise 42:} Prepare a beam line with eight FODO cells without bending magnets
  and with 60 degrees/cell phase advance in both planes. (a) Prepare the periodic beam 
  matrix sigma0 (4x4, uncoupled) as the initial beam and plot both beam sizes along the 
  beam line. (b) Use sigma0 as the starting beam, but change the focal length of the 
  second quadrupole by 10\,\% and plot the beam sizes once again. Discuss you observations.
\item{\bf Exercise 43:} From the lecture about betatron coupling identify the transfer 
  matrix for a solenoid and write a function that receives the longitudinal magnetic
  field $B_s$ and the length of the solenoid as input and returns the transfer matrix. 
  Then extend the simulation code to handle solenoids. Finally, define a beam line where 
  you place the solenoid in the middle of a FODO cell and follow a particle with initial 
  condition $(x_0,x'_0,y_0,y'_0)=(10^{-3}\,\mathrm{m},0,0,0).$ What do you observe? Is the motion 
  confined to the horizontal plane? 
\end{itemize}
And this brings us to the end of this tutorial. Here, we try to give some idea about
what goes on ``under the hood'' of full-blown beam optics codes, such as MADX~\cite{MADX}.
Moreover we hope to have provided some guidance on how to write such a simulation code 
yourself and how to use it to ``play around'' with simple beam-optical system and gets 
a feeling for their behavior. But the story does not stop here. There are many more topic 
covered in the CERN Accelerator School and the more advanced 
textbooks~\cite{VZAPB,WOLSKI,WIEDEMANN,SYLEE}.
%
%
%
%
%
\bibliographystyle{plain}

%
%
\appendix
\section{Propagating the moments of a distribution}
\label{sec:propmom}
Here we work out how the moments propagate through the beam line as a function
of the transfer matrix $R$ that maps particle coordinates from one place to another.
If we assume that the downstream coordinates $\hat x$ are given by $\vec{\hat x} = R\vec x$
with $\vec{\hat x}=(\hat x,\hat x')^t=(\hat x_1,\hat x_2)^t$ and the upstream 
coordinates by $\vec x=(x,x')^t=(x_1,x_2)^t.$ Note that here subscript $1$ labels 
the position and $2$ the angle. The equation that relates $\vec{\hat x}$ to $\vec x$ 
can then be written as $\hat x_i = \sum_{j=1}^2 R_{ij}x_j,$ which is valid for every particle
in the ensemble. We can therefore calculate the averages over the final distribution as
\begin{equation}\label{eq:xtransp}
\hat X_i = \langle \hat x_i\rangle = \langle \sum_{j=1}^2 R_{ij} x_j\rangle 
=  \sum_{j=1}^2 R_{ij}\langle x_j\rangle = \sum_{j=1}^2 R_{ij} X_j\ .
\end{equation}
Here the first equality is simply the definition of $\hat X_i$ and the second follows from
replacing the particle coordinates $\hat x_i$ through the transfer matrix elements $R_{ij}$ 
and the upstream coordinates $x_j.$ Since the same matrix $R$ applies to {\em all} particles,
we can pull it out of the averaging, likewise with the summation over $j.$ But then we are 
left with the averages of $\langle x_j\rangle,$ which is the same as $X_j.$ This simple 
calculation proves the remarkable fact that the {\em beam centroids propagate in the same 
way as individual particles} and we can visualize the beam centroid as some sort of
super-particle. Note, however, that this feature depends crucially that the propagation
is described by a linear operation---a matrix multiplication. If non-linear elements, such
as sextupoles are present in the beam line, this correspondence is no longer strictly true!
\par
Now we consider the second moments, but assume that the centroids are zero $X_i=0,$ which
simplifies the notation. The sigma-matrix elements $\hat\sigma_{ij}$ ``on the other end of 
the transfer matrix $R$'' can be calculated in the following way
\begin{equation}\label{eq:sigma1}
\hat\sigma_{ij} 
  =  \langle \sum_{k=1}^2 R_{ik} x_k \sum_{l=1}^2 R_{jl} x_l \rangle
  =   \sum_{k=1}^2 R_{ik}  \sum_{l=1}^2 R_{jl} \langle x_k x_l \rangle
  =  \sum_{k=1}^2\sum_{l=1}^2 R_{ik} R_{jl} \sigma_{kl}\ ,
\end{equation}
where the first equality follows from inserting $\hat x_i=\sum_{j=1}^2 R_{ij}x_j$ in the
definition of $\hat\sigma_{ij}= \langle \hat x_i \hat x_j \rangle$. Since the transfer 
matrices are constant and summing is linear, we can extract both sum and $R$ from the
averaging, denoted by the angle brackets and find that the sigma matrix transform 
according to
\begin{equation}\label{eq:sigma}
\hat\sigma = R \sigma R^t\ ,
\end{equation}
where we converted the expression from Equation~\ref{eq:sigma1}, which is given in 
components, into a matrix equation. Note that $R^t$ denotes the transpose of the matrix 
$R.$ Equations~\ref{eq:xtransp} and~\ref{eq:sigma} are given in Equation~\ref{eq:prop}
in Section~5 in the main part of the text.
\section{Octave/MATLAB code}
\label{sec:code}
\subsection{\tt beamoptics.m}
This script is used in Section~3 to plot the trajectory of a particle. It is already
explained in the main part of the text.
\begin{verbatim}
% beamoptics.m, 
clear all; close all
D=@(L)[1, L; 0, 1];
Q=@(F)[1, 0; -1/F, 1];
F=2;   % focal length of the quadrupoles
fodo=[ 1,  5,  0.2,  0;    % 5* D(L/10)
       2,  1,  0.0, -F;    % QD
       1, 10,  0.2,  0;    % 10* D(L/10)  
       2,  1,  0.0,  F;    % QF/2
       1,  5,  0.2,  0];   % 5* D(L/10)  
%beamline=fodo;               % name must be 'beamline' 
%beamline=[fodo;fodo];
beamline=repmat(fodo,100,1);
nlines=size(beamline,1);      % number of lines in beamline
nmat=sum(beamline(:,2))+1;    % sum over repeat-count in column 2
Racc=zeros(2,2,nmat);         % matrices from start to element-end
Racc(:,:,1)=eye(2);           % initialize first with unit matrix
spos=zeros(nmat,1);           % longitudinal position
ic=1;                         % element counter
for line=1:nlines             % loop over input elements
  for seg=1:beamline(line,2)  % loop over repeat-count 
     ic=ic+1;                 % next element          
     Rcurr=eye(2);            % matrix in next element
     switch beamline(line,1)  
       case 1   % drift
          Rcurr=D(beamline(line,3));
       case 2   % thin quadrupole
          Rcurr=Q(beamline(line,4));   
       otherwise
          disp('unsupported code')
     end		  
     Racc(:,:,ic)=Rcurr*Racc(:,:,ic-1);    % concatenate 
     spos(ic)=spos(ic-1)+beamline(line,3); % position of element   
  end
end
x0=[0.001;0];         % 1 mm offset at start
data=zeros(1,nmat);   % allocate memory
for k=1:nmat          
  x=Racc(:,:,k)*x0;
  data(k)=x(1);       % store the position
end
plot(spos,1e3*data,'k','LineWidth',2);
xlabel('s [m]'); ylabel(' x [mm]'); xlim([spos(1),spos(end)])
\end{verbatim}
\subsection{\tt calcmat.m}
The function {\tt calcmat()} encapsulates the allocating of matrices and the
calculation of the transfer matrices {\tt Racc} in order to avoid rewriting
this section over and over. The function receives the {\tt beamline} as input
and returns the matrices {\tt Racc}, the positions of the elements {\tt spos}, 
the number of elements (including all segments), and the number of lines in
the {\tt beamline.} 
\begin{verbatim}
  % calcmat.m, calculate the transfer matrices
  function [Racc,spos,nmat,nlines]=calcmat(beamline)
  D=@(L)[1, L; 0, 1];
  Q=@(F)[1, 0; -1/F, 1];
  nlines=size(beamline,1);      % number of lines in beamline
  nmat=sum(beamline(:,2))+1;    % sum over repeat-count in column 2
  Racc=zeros(2,2,nmat);         % matrices from start to element-end
  Racc(:,:,1)=eye(2);           % initialize first with unit matrix
  spos=zeros(nmat,1);           % longitudinal position
  ic=1;                         % element counter
  for line=1:nlines             % loop over input elements
    for seg=1:beamline(line,2)  % loop over repeat-count 
      ic=ic+1;                 % next element          
      Rcurr=eye(2);            % matrix in next element
      switch beamline(line,1)  
      case 1   % drift
        Rcurr=D(beamline(line,3));
      case 2   % thin quadrupole
        Rcurr=Q(beamline(line,4));   
      otherwise
        disp('unsupported code')
      end		  
      Racc(:,:,ic)=Rcurr*Racc(:,:,ic-1);    % concatenate 
      spos(ic)=spos(ic-1)+beamline(line,3); % position of element   
    end
  end
\end{verbatim}
\subsection{\tt R2beta.m}
This function encodes Equation~\ref{eq:betaR}. It receives a transfer matrix $R,$
which is assumed to describe a periodic system, and returns the ``tune'' $Q$ or the
phase advance $\mu$ divided by $2\pi$, and the Twiss parameters $\alpha,\beta,$ and $\gamma.$
\begin{verbatim}
  function [Q,alpha,beta,gamma]=R2beta(R)
  mu=acos(0.5*(R(1,1)+R(2,2)));
  if (R(1,2)<0) mu=2*pi-mu; end
  Q=mu/(2*pi);
  beta=R(1,2)/sin(mu);
  alpha=(0.5*(R(1,1)-R(2,2)))/sin(mu);
  gamma=(1+alpha^2)/beta;
\end{verbatim}
\subsection{\tt propagate\_beam}
This function is used to produce the plots in Figure~\ref{fig:beam},\ref{fig:sigvzs},
and~\ref{fig:sig2}. 
\begin{verbatim}
  % propagate_beam.m
  clear all; close all;
  % pkg load statistics     % only needed for hist3() in octave 
  beamoptics;    % with beamline=repmat(fodo,5,1)
  Npart=10000;
  beam=randn(2,Npart);
  x0=0; sigx=1;     % initial average position and width
  xp0=1; sigxp=0.5; % initial angle and divergence
  beam(1,:)=sigx*beam(1,:)+x0;
  beam(2,:)=sigxp*beam(2,:)+xp0;
  xbins=-4*sigx:0.5*sigx:4*sigx;      % binning for histogram
  %ybins=-4*sigxp:0.5*sigxp:4*sigxp;
  ybins=xbins;
  beam0=beam;   % save for posterity
  show_beam(beam0,xbins,ybins);    % Figure 1 left
  pause(1); figure
  R=[1,1;0,1]; beam=R*beam0; 
  show_beam(beam,xbins,ybins);     % Figure 1 right
  figure                           % Figure 2+3 super-imposed
  data=zeros(nmat,2);
  for k=1:nmat                     % propagate particles
    beam=Racc(:,:,k)*beam0;    
    data(k,1)=std(beam(1,:));      % data for Figure 2
  end
  X0=[x0;xp0];                     % initial centroids
  sigma0=[sigx^2,0;0,sigxp^2];     % and sigma0
  for k=1:nmat
    X=Racc(:,:,k)*X0;              % Equation 7
    sigma=Racc(:,:,k)*sigma0*Racc(:,:,k)';
    data(k,2)=sqrt(sigma(1,1));
  end
  plot(spos,data(:,1),'k',spos,data(:,2),'*')
  xlabel('s [m]'); ylabel('\sigma_x [mm]')
\end{verbatim}
Note that in octave we have to load the {\tt statistics} package with the
command {\tt pkg load statistics}, which is needed for {\tt hist3()} to
prepare the three-dimensional histograms in {\tt show\_beam()}. The 
package is a available from \url{https://octave.sourceforge.io/}. It
needs to be installed first, but only once.
\subsection{\tt show\_beam}
This function is used to display the beam in Figure~\ref{fig:beam}.
\begin{verbatim}
  % show_beam.m
  function show_beam(beam,xbins,ybins)
  subplot(2,2,1); 
  contour(xbins,ybins,hist3(beam',{xbins,ybins})');
  xlabel('x [mm]'); ylabel('x'' [mrad]');
  subplot(2,2,2); 
  hist(beam(2,:),ybins);  xlabel('x'' [rad]'); camroll(90)
  title(['Mean = ',num2str(mean(beam(2,:)),2), ...
    ' \pm ',num2str(std(beam(2,:)),2)])
  subplot(2,2,3); 
  hist(beam(1,:),xbins); xlabel('x [m]');
  title(['Mean = ',num2str(mean(beam(1,:)),2), ...
    ' \pm ',num2str(std(beam(1,:)),2)])
  subplot(2,2,4); 
  hist3(beam',{xbins,ybins}); 
  xlabel('x [mm]'); ylabel('x'' [mrad]');
\end{verbatim} 

%
%
\end{document}